\input harvmac.tex
\input epsf
\noblackbox
\def\figin{\epsfcheck\figin}\def\figins{\epsfcheck\figins}
\def\epsfcheck{\ifx\epsfbox\UnDeFiNeD
\message{(NO epsf.tex, FIGURES WILL BE IGNORED)}
\gdef\figin##1{\vskip2in}\gdef\figins##1{\hskip.5in}
\else\message{(FIGURES WILL BE INCLUDED)}%
\gdef\figin##1{##1}\gdef\figins##1{##1}\fi}
\def\DefWarn#1{}
\def\figinsert{\goodbreak\midinsert}
\def\ifig#1#2#3{\DefWarn#1\xdef#1{fig.~\the\figno}
\writedef{#1\leftbracket fig.\noexpand~\the\figno}%
\figinsert\figin{\centerline{#3}}\medskip\centerline{\vbox{\baselineskip12pt
\advance\hsize by -1truein\noindent\footnotefont{\bf Fig.~\the\figno } \it#2}}
\bigskipw\endinsert\global\advance\figno by1}


\def\encadremath#1{\vbox{\hrule\hbox{\vrule\kern8pt\vbox{\kern8pt
 \hbox{$\displaystyle #1$}\kern8pt}
 \kern8pt\vrule}\hrule}}
 %
 %
 

 \font\cmss=cmss10
 \font\cmsss=cmss10 at 7pt
 \def\rlx{\relax\leavevmode}
 \def\inbar{\vrule height1.5ex width.4pt depth0pt}
 \def\IC{\relax\,\hbox{$\inbar\kern-.3em{\rm C}$}}
 \def\IN{\relax{\rm I\kern-.18em N}}
 \def\IP{\relax{\rm I\kern-.18em P}}

\def\ZZ{\rlx\leavevmode\ifmmode\mathchoice{\hbox{\cmss Z\kern-.4em Z}}
  {\hbox{\cmss Z\kern-.4em Z}}{\lower.9pt\hbox{\cmsss Z\kern-.36em Z}}
  {\lower1.2pt\hbox{\cmsss Z\kern-.36em Z}}\else{\cmss Z\kern-.4em Z}\fi}
 \def\IZ{\relax\ifmmode\mathchoice
 {\hbox{\cmss Z\kern-.4em Z}}{\hbox{\cmss Z\kern-.4em Z}}
 {\lower.9pt\hbox{\cmsss Z\kern-.4em Z}}
 {\lower1.2pt\hbox{\cmsss Z\kern-.4em Z}}\else{\cmss Z\kern-.4em Z}\fi}
 \def\IZ{\relax\ifmmode\mathchoice
 {\hbox{\cmss Z\kern-.4em Z}}{\hbox{\cmss Z\kern-.4em Z}}
 {\lower.9pt\hbox{\cmsss Z\kern-.4em Z}}
 {\lower1.2pt\hbox{\cmsss Z\kern-.4em Z}}\else{\cmss Z\kern-.4em Z}\fi}

 \def\narrowplus{\kern -.04truein + \kern -.03truein}
 \def\narrowminus{- \kern -.04truein}
 \def\narrowminussub{\kern -.02truein - \kern -.01truein}

 \def\frac#1#2{{#1\over #2}}

 \def\IZ{\relax\ifmmode\mathchoice
 {\hbox{\cmss Z\kern-.4em Z}}{\hbox{\cmss Z\kern-.4em Z}}
 {\lower.9pt\hbox{\cmsss Z\kern-.4em Z}}
 {\lower1.2pt\hbox{\cmsss Z\kern-.4em Z}}\else{\cmss Z\kern-.4em Z}\fi}
 \def\IB{\relax{\rm I\kern-.18em B}}
 \def\IC{{\relax\hbox{$\inbar\kern-.3em{\rm C}$}}}
 \def\Ic{{\relax\hbox{$\inbar\kern-.22em{\rm c}$}}}
 \def\ID{\relax{\rm I\kern-.18em D}}
 \def\IE{\relax{\rm I\kern-.18em E}}
 \def\IF{\relax{\rm I\kern-.18em F}}
 \def\IG{\relax\hbox{$\inbar\kern-.3em{\rm G}$}}
 \def\IGa{\relax\hbox{${\rm I}\kern-.18em\Gamma$}}
 \def\IH{\relax{\rm I\kern-.18em H}}
 \def\II{\relax{\rm I\kern-.18em I}}
 \def\IK{\relax{\rm I\kern-.18em K}}
 \def\IP{\relax{\rm I\kern-.18em P}}

 \font\cmss=cmss10 \font\cmsss=cmss10 at 7pt
 \def\IR{\relax{\rm I\kern-.18em R}}

 %

 %
 %
 \def\eqnn#1{\xdef
#1{(\secsym\the\meqno)}\writedef{#1\leftbracket#1}%
 \global\advance\meqno by1\wrlabeL#1}
 \def\eqna#1{\xdef
#1##1{\hbox{$(\secsym\the\meqno##1)$}}

\writedef{#1\numbersign1\leftbracket#1{\numbersign1}}%
 \global\advance\meqno by1\wrlabeL{#1$\{\}$}}
 \def\eqn#1#2{\xdef
#1{(\secsym\the\meqno)}\writedef{#1\leftbracket#1}%
 \global\advance\meqno by1$$#2\eqno#1\eqlabeL#1$$}

\newdimen\tableauside\tableauside=1.0ex
\newdimen\tableaurule\tableaurule=0.4pt
\newdimen\tableaustep
\def\phantomhrule#1{\hbox{\vbox to0pt{\hrule height\tableaurule width#1\vss}}}
\def\phantomvrule#1{\vbox{\hbox to0pt{\vrule width\tableaurule height#1\hss}}}
\def\sqr{\vbox{%
  \phantomhrule\tableaustep
  \hbox{\phantomvrule\tableaustep\kern\tableaustep\phantomvrule\tableaustep}%
  \hbox{\vbox{\phantomhrule\tableauside}\kern-\tableaurule}}}
\def\squares#1{\hbox{\count0=#1\noindent\loop\sqr
  \advance\count0 by-1 \ifnum\count0>0\repeat}}
\def\tableau#1{\vcenter{\offinterlineskip
  \tableaustep=\tableauside\advance\tableaustep by-\tableaurule
  \kern\normallineskip\hbox
    {\kern\normallineskip\vbox
      {\gettableau#1 0 }%
     \kern\normallineskip\kern\tableaurule}%
  \kern\normallineskip\kern\tableaurule}}
\def\gettableau#1 {\ifnum#1=0\let\next=\null\else
  \squares{#1}\let\next=\gettableau\fi\next}

\tableauside=1.0ex
\tableaurule=0.4pt

\def\IE{\relax{\rm I\kern-.18em E}}
\def\IP{\relax{\rm I\kern-.18em P}}

\Title
{\vbox{
 \baselineskip12pt
\hbox{HUTP-05/A043}}}
 {\vbox{
 \centerline{The String Landscape and the Swampland}
 \centerline{}
 }}
\centerline{Cumrun Vafa}

\bigskip\centerline{ Jefferson Physical Laboratory, Harvard University}
\centerline{Cambridge, MA 02138, USA}\smallskip

\smallskip
 \vskip .3in \centerline{\bf Abstract}

{Recent developments in string theory suggest that string theory
landscape of vacua is vast.  It is natural to ask
if this landscape is as vast as allowed by consistent-looking
effective field theories.  We use universality ideas from string theory
to suggest that this is not the case, and that the landscape is surrounded by an even more vast
swampland of
consistent-looking semiclassical effective
field theories, which are actually inconsistent.  Identification of the 
boundary of the landscape is a central
question which is at the heart of the meaning of universality
properties of consistent quantum gravitational theories.
We propose certain finiteness criteria as one relevant factor
in identifying this boundary
(based on talks given at the Einstein Symposium in Alexandria, at the
2005 Simons Workshop in Mathematics and Physics, and the talk
to have been presented at Strings 2005). }

 \smallskip \Date{September 2005}

\newsec{Introduction}

The central problem in string theory is how it will eventually
connect with experiments.  In this regard the notion of what kind
of field theories can arise in string theory becomes a central question.
The diversity of string vacuum constructions, which has been exploding in
the past few years, leads one to the picture that basically any effective
field theory which looks at least semiclassically consistent can arise
in string theory.    This is in sharp contrast, but not in contradiction,
with the fact that the theory itself is believed to be unique.

If indeed any consistent looking effective field theory is actually
consistent,  one would naturally
wonder about the wisdom of trying to construct string vacua using
complicated geometries of internal manifolds etc.  Instead one could just
consider an effective field theory which looks consistent and which most
closely resembles experimental data.  In this context it is natural to postpone the question of its
stringy construction until we probe high enough energies where the
effective field theory description breaks down and a more stringy approach
is called for.  If consistency of a quantum theory of gravity does
not pick out which effective field theories can arise, then string theory
becomes of far less relevance for questions of IR physics.

In this note we argue that the pendulum has swung far enough:  Not all
consistent looking effective field theories are actually consistent and
one should not be misled by a vast array of string vacuum constructions; 
we argue that 
the consistent looking effective field theories which are actually inconsistent
are even more vast.  This vast series of semiclassically consistent effective
field theories which are actually inconsistent we shall call the `swampland'.   Thus we
view the vast stringy landscape of vacua as a relatively small island in an even more vast swampland
of quantum inconsistent but semiclassically consistent effective field theories. 

To argue the above picture we appeal to the fact that certain general
features emerge in {\it all} string theory\foot{In this paper
whenever we refer to string theory we have in mind all possible
vacuum constructions including those which are more naturally
phrased in terms of M-theory or F-theory.} vacuum constructions,
which have no apparent explanation as consistency conditions for effective field theories.
Even though this by itself does not {\it prove} the above picture, it makes it
look rather plausible.  Moreover in some sense {\it most} of the consistent
looking effective theories turn out to fall in the swampland\foot{In fact
in the most naive sense of measure, the string landscape is of measure zero
compared to the swampland.}.

The restrictions we focus on are finiteness properties including finiteness
of volume of scalar fields, finiteness of the number of fields and finiteness/restrictions
on the rank of the gauge groups.   These finiteness properties can be relaxed
if we turn off the gravity.  In other words they are {\it directly correlated
with having a consistent quantum gravity coupled to matter}.  
As it turns out these finiteness issues are highly non-trivial and are deeply related to the dualities
in string theory and in fact {\it imply} the existence of certain S-dualities.  We view
these finiteness crieteria as a universal feature of any potentially consistent
quantum theory of gravity coupled to matter.

Some aspects of finiteness properties of string landscape have been independently
noted by Michael Douglas \ref\md{M. Douglas, Talk given in Strings 2005, Toronto.}.
The main motivation to consider the finiteness issues there was the requirement
of being able to give a notion
of a number distribution on the stringy landscape.

\newsec{Finiteness of Scalar Field Moduli Space}
Consider a quantum theory of gravity, coupled to massless scalar fields
$\Phi^i$
with no potential 
$$V(\Phi^i)=0.$$
There could also be some other fields in the theory, but we will
concentrate on the massless fields.
We have the low energy effective action
$$S=\int {M_P}^{d-2} R+ g_{ij}(\Phi)\partial \Phi^i \partial \Phi^j+... $$
where $...$ represents higher derivative terms, as well as possibly other
fields in the theory.  Consider the volume of the scalar
field space
$$V_{\Phi}=\int d\Phi \sqrt{g(\Phi)}$$
Let us be a bit more precise:  As we vary the value of $\Phi$
the effective field theory may break down, due to the appearance
of light states.  Let us treat the effective field theory as being
valid for energy scales
$$E\ll \Lambda$$
where $\Lambda$ is some fixed scale.  Let us consider regions
in $\Phi$ space where all the integrated out massive excitations have masses
$m\gg \Lambda$.  Let us call this region $B^\Lambda$.  Consider the refined volume
in the field space
$$V_{\Phi}^\Lambda=\int_{B^\Lambda} d\Phi \sqrt{g(\Phi)}$$
\lref\Dimo{S.~Dimopoulos, S.~Kachru, J.~McGreevy and J.~G.~Wacker,  ``N-flation,'' 
arXiv:hep-th/0507205.}

We define
$$V_\Phi=V_\Phi^{\Lambda}\big|_{\Lambda\rightarrow 0}$$
and view $\Lambda$ as a regulator.

From the viewpoint of effective field theory there
appears to be nothing inconsistent with 
$$V_{\Phi} =\infty.$$
For example consider the case where we have one real field $\Phi$
taking value in ${\bf R}$ and where no masses depend on the value of $\Phi$.
So the region $B$ is the entire real line and so $V_{\Phi} =\infty$
in this case.
However, as it turns out in {\it all} examples which come
from string theory this is indeed finite\foot{In the context of volume of field space for certain
axion fields this finiteness
was studied recently \Dimo.}
 as long as the Newton's constant
is not zero (i.e. $M_p\not= \infty$ and  gravity is not decoupled):
$$V_{\Phi} \not=\infty.$$
(more precisely the conjecture is that this could go to infinity $\sim |{\rm log} \Lambda |$ 
as $\Lambda \rightarrow 0$ but no
worse.)
Let us give examples of this principle from string theory
and also discuss how it can be evaded {\it if} we decouple gravity. 

Consider type IIB in 10 dimensions.  This theory has a complex
scalar $\tau$, which is the combination of string coupling constant combined with a scalar
RR field.  The conjecture applied to this case states that
$$V_\tau ={\rm finite}$$

This is given by 
$$V_\tau= \int {d^2\tau\over \tau_2^2}$$
If we did not use S-duality of type IIB this would indeed be
infinite, violating the above conjecture.  Thus we see that
the finiteness of the scalar volume in this case is closely
related to the {\it existence} of some non-trivial S-dualities.  With S-duality
taken into account this volume is indeed finite.

As another series of example
consider compactifications of type II strings on $T^d$. The salar
fields are parameterized by the Narain moduli space
$$SO(d,d)/SO(d)\times SO(d)\times SO(d,d;{\bf Z})$$
which is known to be finite \ref\jka{A. Borel and Harish-Chandra,
``Arithmetic Subgroups of Algebraic Groups,''  Ann. of Math. {\bf 75}, 485 (1962).}\foot{For $d=1$ finiteness
follows once we consider the cutoff dependent $V_{\Phi}^\Lambda$ which
behaves as $|{\rm log} \Lambda |$.}.  The existence
of T-duality is {\it crucial} in making this volume
finite.

\lref\HorneMI{
  J.~H.~Horne and G.~W.~Moore,
  ``Chaotic coupling constants,''
  Nucl.\ Phys.\ B {\bf 432}, 109 (1994)
  [arXiv:hep-th/9403058].
}
\lref\TodorovJG{
  A.~Todorov,
  ``Weil-Petersson volumes of the moduli spaces of CY manifolds,''
  arXiv:hep-th/0408033.
}

Similarly if we consider compactifications of type II
strings on $K3$, the scalar field space is given by
$$ SO(20,4)/SO(20)\times SO(4)\times SO(20,4;{\bf Z})$$
which is finite, again thanks to the non-trivial
T-dualities leading to quotient by $SO(20,4;{\bf Z})$ in the
above.  Similarly one can consider compactifications
of type II on Calabi-Yau manifolds.  The vector multiplet
have scalars which parameterize moduli space of complex
structure (or quantum corrected K\"{a}hler structure) for type IIB (type
IIA). These are believed to be finite in volume (see discussion
in \refs{\HorneMI , \TodorovJG} ).

Note that this finiteness property correlates with having a finte
Newton's constant in the gravitational sector.  In particular
if we consider decompactifying the internal geometry,
which leads to the lower dimensional gravitational sector
to be non-dynamical, with vanishing Newton's constant,
the volume of scalar field space is no longer finite.
For example consider type II strings on non-compact toric
Calabi-Yau manifolds, which leads to
vanishing Newton's constant in the lower dimensional
theory due to the fact that the internal volume is infinite.  This leads to the volume of the 
 scalar moduli space, i.e. moduli space of
non-compact toric Calabi-Yau not being finite either (this is a well known
fact in the context of geometric engineering, where the volume of the
non-compact moduli space gets mapped to the volume of the space for scalar fields
in the ${\cal N}=2$ supersymmetric gauge theories, which is clearly infinite).

This finiteness can be rephrased in the context
of AdS/CFT in the following way\foot{This point was
developed in discussions with E. Silverstein.}. The moduli
of scalar fields in the AdS context gets translated to the moduli
of the CFT.  Thus the finiteness of the moduli of scalar
fields is the statement, in the CFT context, of the finiteness
of the moduli of the CFT.  Finiteness of Newton's constant
in the AdS context correlates with the existence of discrete
spectrum of operators in the CFT (rather than a continuous
spectrum).  This is thus a purely field theoretic restatement
of the gravitational finiteness type questions we are referring to:
\vglue .5cm
{\it Is the volume of moduli space of CFT's, with a discrete
spectrum of operators, finite}?
\vglue .5cm
In all the examples we know of, for example superconformal field
theories in 4 dimensions, this seems to be true.  It also seems to be
true for superconformal field theories in 2 dimensions. This it turns
out gets related to the finiteness of volume of moduli space of scalar fields parameterizing
internal geometry,
in the context of string theory:  In that case the worldsheet CFT moduli space
is the same as the moduli space of the corresponding scalar fields parameterizing
internal geometry.  So
this is at least a consistency check on this conjecture.

\newsec{Finiteness of Matter Fields}
Consider compactifying string theory on a manifold.  The massless
modes of the compactified theory, apart from the gravity multiplet,
will have a number of other fields which get related to certain
cohomology classes, or certain cohomology computations on the manifold.
The number of such matter fields is thus directly bounded by the dimension
of the relevant cohomology of the manifold.  In all the examples we know
in string theory, the dimension of cohomology of the manifold is finite.
For example, we do not know of a sequence of Calabi-Yau threefolds for which
the dimension of cohomologies can get unboundedly large.  It is not
known if this is actually a true mathematical fact, but
experience with various examples in string theory gives one a strong
feeling that this is going to be finite.  

The interesting point is that again this finiteness correlates with the
Newton's constant being non-zero:  If we consider local geometries for
`compactification' which do not have finite volume, then the dimension
of the matter fields can grow arbitrarily large:  Consider for example
gauge theories in 6 dimensions.  If we consider type IIA in the presence
of $A_{N-1}$ singularities, we obtain a $U(N)$ gauge theory.  There
is no bound on $N$ and so we can have an arbitrarily large rank gauge group,
in accordance with field theoretic semiclassical consistency conditions.  But
if we wish to get a non-trivial gravity in 6 dimensions by having a non-zero
Newton's contant then we need to embed this in a compact geometry and it is
known that this cannot be done for arbitrarily large $N$ in $K3$ (indeed
the rank of the matter field gauge group would be bounded by 20).  Thus the 
consistency conditions for an effective {\it gravitational} theory should be more subtle than
the corresponding condition for a consistent gauge theory which can be constructed
in string theory for arbitrary rank.

\subsec{A Lower Bound Conjecture}
In the above discussion we have considered upper bounds
on the number of fields.
  In
this section we discuss the opposite restriction, namely that in some
theories there should be {\it some} scalar fields, which is not demanded
by naive consistency of the effective field theory.  This again is
going to be hard to prove.  But let us present one potential candidate:
Consider ${\cal N}=2$ supersymmetric theories in 4d and ask if there are
any consistent theories which have, as the low energy limit,
 only the supergravity multiplet and nothing
else.  Recall that the pure ${\cal N}=2$ supergravity multiplet has no scalar.
There is no known construction within string theory which accomplishes this.
All known examples have extra scalars, such as the string coupling constant
or the volume of the internal manifold etc.  It is natural to conjecture
that there is at least one such scalar in any such theory and that pure ${\cal N}=2$
supergravity does not arise as the effective theory of a fully consistent quantum
theory.  It would be interesting to verify or disprove this.

\newsec{Restrictions on Gauge Fields}

The gauge fields that arise in string theory seem more restricted
than would have been expected based on the requirement
of semiclassical consistency of the effective field theory.
There are many such examples.  For example consider $N=1$ supersymmetric
theories in 10 dimensions with gauge groups:
$$ U(1)^{496}; U(1)^{248}\times E_8 $$
Both of these theories are consistent as far as anomalies are concerned
\ref\gs{M.~B.~Green and J.~H.~Schwarz,
``Anomaly Cancellation In Supersymmetric D=10 Gauge Theory And Superstring
Theory,''
  Phys.\ Lett.\ B {\bf 149}, 117 (1984).}.
Thus they both appear to be consistent semiclassically.
However, we have no way of constructing either of these theories in string theory.
It is hard to believe that they actually exist, though we have no proof of this.  It would
be very interesting either to construct these or what is more likely, to find
what consistency condition goes wrong with these theories which renders them inconsistent\foot{
One proposal along these lines, suggested by M. Douglas, is to assume the existence
of the 1-brane electrically charged under the anti-symmetric tensor field and study the consistency
conditions of the theory living on it.  It would be interesting to see if this
can be completed to an argument ruling out the existence of these theories.}

There are many more such examples.  For example if one considers
consistent supersymmetric gauge theories coupled to gravity in 9
dimensions, there is a very restricted list known \ref\gaugth{ J.~de Boer, R.~Dijkgraaf, K.~Hori, A.~Keurentjes, J.~Morgan, D.~R.~Morrison and S.~Sethi,
  Adv.\ Theor.\ Math.\ Phys.\  {\bf 4}, 995 (2002)
  [arXiv:hep-th/0103170].}\ and it is
difficult to imagine there are many other constructions within string theory.  
The importance of finding these restrictions for a small dimension of internal
manifold is that
it would be easier to be convinced that there are no other constructions
missing.  Were we to go further down in dimensions and in supersymmetry
the available possibilities for compactifications, including the geometry
and the flux/brane data would be so enormous that it would have been
harder to make a definitive conclusion.  Keeping the question
at higher dimensions makes it essentially clear that there is no
construction within string theory which gives other results.

\newsec{Concluding Thoughts}

We feel that finding suitable techniques to answer the types of questions raised in this short note
are critical in the next level of our understanding of string theory and what it means to have
a consistent quantum theory of gravity coupled to matter.  We believe
string theory can serve as a testing ground for any proposed consistency
condition.  The conjectures motivated in this note are motivated from
the constructability or lack thereof within string theory.  Examples
of such potential questions include the interesting work \ref\dha{ A.~Dabholkar and J.~A.~Harvey,
``String islands,'' JHEP {\bf 9902}, 006 (1999)
  [arXiv:hep-th/9809122].}\
which seeks to answer which minimal supergravities with at least one scalar field
 can be realized in string
theory.
It
would be very interesting to find some general patterns of what
are not possible to construct within string theory but that
are naively allowed as consistent anomaly free effective theories (for some
other possible constraints see \ref\nimlu{N. Arkani-Hamed, L. Motl, A. Nicolis and
C. Vafa, work in progress.}).
There clearly should be many such criteria besides the ones discussed here.
It would be exciting to uncover these conditions and come up with what it
means to talk about the universality conditions for consistent quantum theories
of gravity.

\vglue 2cm
\bigskip
\noindent {\bf Acknowledgments}
\bigskip
\noindent
I would like to thank M. Aganagic,
N. Arkani-Hamed, R. Dijkgraaf, M. Douglas, D. Kazhdan, L. Motl, N. Shadbeh, E. Silverstein,
D. Sullivan and E. Witten for valuable discussions.  I have greatly benefited
from participation in the Einstein Symposium in Alexandria and the
2005 Simons Workshop in Mathematics and Physics
where the above work was presented.

This research is supported in part
by NSF grants PHY-0244821 and DMS-0244464.
\listrefs
\end
\bye